\documentclass[a4paper,11pt]{article}
\usepackage{pos}
\usepackage{caption}
\usepackage{subcaption}

\title{\texttt{tilepy}: Smart Scheduling for Multi-Messenger Astronomy from Earth to Orbit
}

\author*[a]{Halim Ashkar}
\author[b]{Monica Seglar-Arroyo}
\author[c]{Fabian Schussler}
\author[c]{Weizmann Kiendrébéogo}
\author[d]{Mathieu de Bony de Lavergne}

\affiliation[a]{Institute of Space Sciences (IEEC-CSIC), Campus UAB, Torre C5, 2a planta, 08193 Barcelona, Spain}

\affiliation[b]{IFAE, The Barcelona Institute of Science and Technology, Campus UAB, 08193 Bellaterra (Barcelona), Spain}

\affiliation[c]{IRFU, CEA, Université Paris-Saclay, F-91191 Gif-sur-Yvette, France}

\affiliation[d]{Centre de Physique des Particules de Marseille (CPPM), Aix–Marseille Université, CNRS/IN2P3, 163, Avenue de Luminy – Case 902, 13288 Marseille cedex 9, France}

\emailAdd{astro.tilepy@gmail.com}

\abstract{The rise of direct detection of gravitational waves (GWs) started a new era in multi-messenger astrophysics. Like GWs, many other astrophysical transient sources suffer from poor localization, which can span tens to thousands of square degrees in the sky. Moreover, as the detection horizon for these transients widens and the detection rate increases, current electromagnetic follow-up facilities require tools to optimize the follow-up of poorly localized events and save valuable telescope time for their time-domain astrophysics programs. We present \texttt{tilepy}, a Python library, and a tool to optimize the follow-up of poorly localized transient events. \texttt{tilepy} is used for GWs as well as other poorly localized events such as gamma-ray bursts detected by Fermi-GBM and neutrino candidates from IceCube. \texttt{tilepy} has also been optimized to integrate smoothly with multiple ground-based observatories operating individually or simultaneously with diverse observational configurations. In this contribution, we introduce the latest developments from \texttt{tilepy}, mainly the ability to operate with space-based observatories while taking into consideration factors such as Earth, Sun, and Moon occultation and South Atlantic Anomaly passage. We present innovations to the platform, handling a variety of field of view shapes, the possibility of optimizing observation scheduling with artificial intelligence tools and examples of its use on transient astrophysical events.}

\ConferenceLogo{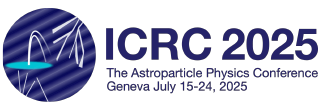}

\FullConference{39th International Cosmic Ray Conference (ICRC2025)\\
 15–24 July 2025\\
Geneva, Switzerland\\}

\begin{document}
\maketitle

\section{Introduction}
GW events suffer from poor localization, which can span tens to thousands of square degrees in the sky. The follow-up of GWs is very challenging, but in the case of an electromagnetic counterpart detection - like in the case of GW170817~\cite{GW170817, GW170817_GRB, GW170O817_MWL} that established neutron star mergers as one of the main sites for particle acceleration and heavy element production in the universe - the source becomes a gold mine for scientists. As the detection horizon for GWs widens (hundreds of Mpc instead of tens of Mpc), the chances of detecting a kilonova in the optical domain are diminishing with current capabilities~\cite{Mochkovitch} given the large localization areas and distances. On the GRB side, the detection rates will rise as well with several small missions planned, such as BurstCube~\cite{Racusin}. Not all GRB detections are well localized in the sky. Sometimes the localization can span several tens of degrees in the sky. This is in particular relevant for the GRBs detected by Fermi-GBM, which is currently the instrument that detects GRBs most frequently. Fermi-GBM GRBs have localization regions with an average 68\% radius of 5.3 degrees. This is similar for the GRBs detected by SVOM-GRM. Although these missions excel in detection rates, the localization of GRBs still presents a challenge. In addition, many other transients suffer from poor localizations, in particular, neutrino candidate detections from IceCube and KM3NET. With the new generation of increasingly sensitive instruments dedicated to time-domain astronomy (e.g. SKAO, LSST, CTAO), knowing how to select the sources of interest becomes crucial. It has become more critical than ever to maximize the efficiency of limited observing time by strategically prioritizing high-probability sky regions, enabling the follow-up of a greater number of events while conserving telescope resources for the most scientifically promising cases.

\texttt{tilepy} is a Python library designed to optimize the follow-up of poorly localized transient events. It does so by maximizing the probability of detecting electromagnetic counterparts, leveraging priors such as skymap probability distributions (2D-strategy) and the spatial distribution of galaxies and possibly their stellar masses within localization regions (3D-strategy). This approach significantly narrows down the search area—from wide sky patches to a small number of targeted galaxies~\cite{Technical_paper}. \texttt{tilepy} also supports coordination across multiple ground-based observatories around the globe with varying observational strategies and preferences, enabling simultaneous observations that enhance sky coverage while minimizing overlap~\cite{tilepy_2}. \texttt{tilepy} provides high level functions that can be directly used for scheduling observations and  medium and low classes and tools that can be directly implemented in telescope response systems. The official website is \url{https://tilepy.com}. tilepy’s code is publicly available on GitHub (\url{https://github.com/astro-transients/tilepy}) and can be used by external users. tilepy can also be accessed directly through its API (available on the website) or through the Astro-COLIBRI platform in one click (\url{https://astro-colibri.science}). \texttt{tilepy} is a working package of the ANR project Multi-messegner Observations of Transients Events (MOTS) project and the Astrophysics Center for Multi-messegner studies in Europe (ACME, WP5). It is integrated within Astro-COLIBRI. It is currently used in H.E.S.S.~\cite{Technical_paper}, CTA~\cite{Patricelli}, and LST~\cite{Carossi}. In this contribution, we present the latest advancements in \texttt{tilepy}, including its new capability to operate with space-based observatories, support for telescopes with all types of fields of view and the integration of novel artificial intelligence tools for optimized observation scheduling.

\begin{figure}[h!]
    \centering
    \includegraphics[width=0.65\textwidth]{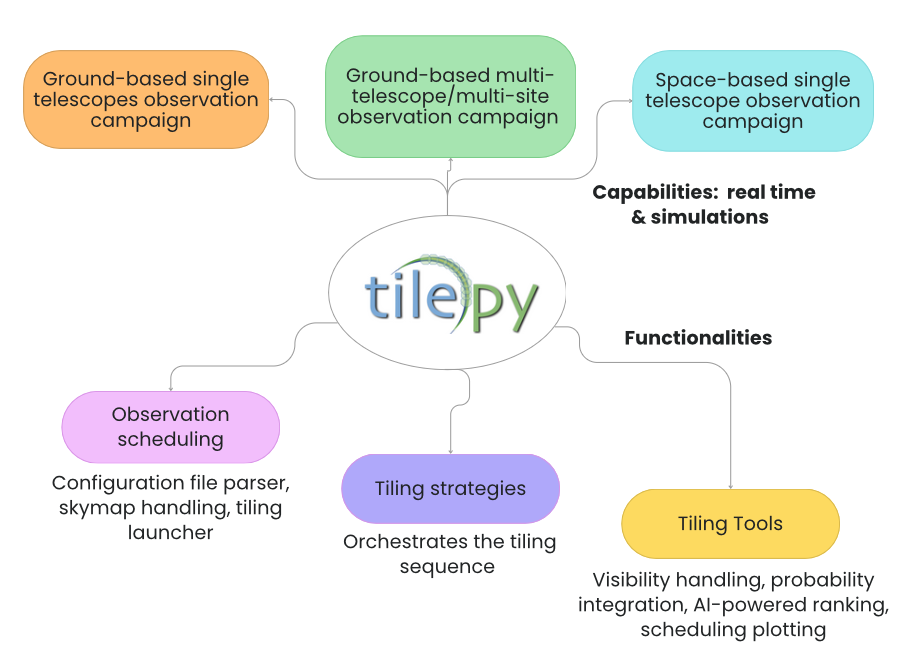}
    \caption{Overview of the Tilepy software.}
    \label{fig:cd}
\end{figure}

\section{Space-Based Scheduling with \texttt{tilepy}}
The observation scheduling with space-based observatories follows a multi-step procedure:

\begin{itemize}
    \item Load the probability sky map, galaxy catalog, and optional exclusion zones (e.g., previously observed regions). In the case of 3D strategies, galaxies within the localization region are assigned probabilities based on their distance, sky position, and stellar mass (see Ashkar et al. for details). the software also reead the user preferences like the start time of observations, the total observations duration adn the number of observations required. 

    \item Apply an appropriate pixelization to the sky map using a specified or automatically optimized \texttt{nside} parameter.

    \item Generate a coordinate grid within the sky localization region. This is typically done by extracting positions within a 90\% confidence contour. This functionality can also be used independently for both ground- and space-based observatories. An example is shown in Figure~1. The software starts by computing test positions within the confidence contour, creating a grid of candidate coordinates. At each position, the probability is evaluated—either by integrating the 2D map probability within the region or, in the case of a 3D strategy, by summing the galaxy-assigned probabilities. The algorithm iterates over all possible positions and selects the top-ranked regions according to the user’s preferences. A minimum probability threshold can also be specified, excluding regions below this limit from the list of primary targets. This approach controls the level of overlap desired in the observations. 
    
    \item Up to this step, the functionality can be used independently of the observatory's location and is applicable to both ground- and space-based observatories.

    \item For space-based observatories, the algorithm connects to external services (specified by the user) to retrieve the spacecraft’s trajectory.

    \item  The algorithm computes the time intervals during which the spacecraft crosses the South Atlantic Anomaly (SAA), a feature helpful for Low Earth Orbit (LEO) satellites. The resulting output is provided to the user and can be used as needed. To identify passages through the South Atlantic Anomaly (SAA), the algorithm calculates the geomagnetic field strength at the scheduled follow-up times. When the magnetic field strength falls below a predefined threshold, indicating a weakened geomagnetic shielding, the spacecraft is flagged as passing through the SAA. This determination relies on coefficients from the International Geomagnetic Reference Field (IGRF) model, specifically the latest 13th generation (IGRF-13~\cite{saa}). The default threshold is set to 25000 nT. 
    
    \item For each specified time interval, the algorithm computes the regions occulted by the Earth, Sun, and Moon. A user-defined safety margin can be applied—for example, a 10-degree exclusion zone around the Sun to protect the detectors. For every time slot, the visibility of each high-probability position (selected in Step~X) is evaluated. Some positions may remain visible throughout the interval, while others might be intermittently obscured. 
    
    \item Two possible output formats are provided to the user: (i) the list of positions that are observable during all specified time intervals, and (ii) the list of positions available at each individual time interval. In addition, the SAA passages of the spacecraft are also provided.
\end{itemize}

\section{Example: Swift-XRT (or UVOT) Scheduling with \texttt{tilepy}}
\label{sec:example}

We present an example of space-based observatory scheduling using the \texttt{tilepy} package. For this demonstration, we consider the gravitational-wave event S250328ae from the O4 observing run, whose 90\% credible region covers an area of 14.2 deg\textsuperscript{2}. We simulate a follow-up campaign using the Swift-XRT observatory. We select  2025-07-28 10:00:10 UTC as a representative start time for observations. The observation campaign is configured as follows:

\begin{itemize}
    \item \textbf{Total duration of observations:} 90 minutes
    \item \textbf{Number of observations:} 30
    \item \textbf{Exposure time per observation:} 3 minutes
    \item \textbf{Instrument constraints (from~\cite{swift_handbook}):}
        \begin{itemize}
            \item Sun-avoidance angle: 47$^\circ$
            \item Moon-avoidance angle: 23$^\circ$ \ (for UVOT)
            \item Earth limb-avoidance angle: 28$^\circ$ \ (conservative margin to avoid atmospheric scattering and noise)
        \end{itemize}
    \item \textbf{Field of view (FoV):} 0.39$^\circ$ \ for both Swift-XRT
\end{itemize}

\begin{figure}[h!]
    \centering
    \begin{subfigure}[t]{0.48\textwidth}
        \centering
        \includegraphics[width=\textwidth]{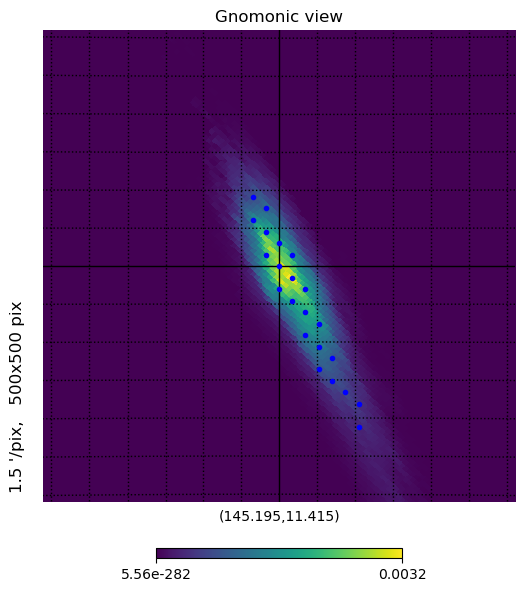}
        \caption{Optimized sky tiling for Swift-XRT using \texttt{tilepy}. The blue dots represent the optimal pointings.}
        \label{fig:tiling}
    \end{subfigure}
    \hfill
    \begin{subfigure}[t]{0.48\textwidth}
        \centering
        \includegraphics[width=\textwidth]{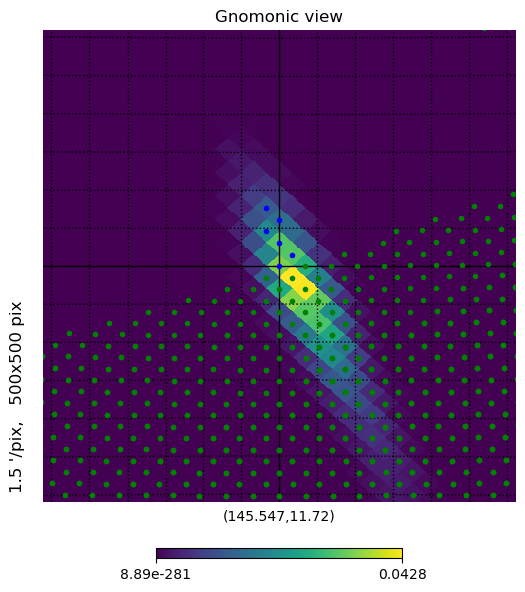}
        \caption{HEALPix pixels occulted by Earth, Sun, or Moon The green  dots represent the occulted positions and the blue dots are the remaining optimal positions after 90 minutes.}
        \label{fig:occulted}
    \end{subfigure}
    \caption{Tiling and occultation maps for the Swift-XRT follow-up of S250328ae.}
    \label{fig:tiling_occulted_combined}
\end{figure}

The optimized tiling strategy produced by \texttt{tilepy} is displayed in Fig.~\ref{fig:tiling}.

\begin{figure}
    \centering
    \includegraphics[width=1\textwidth]{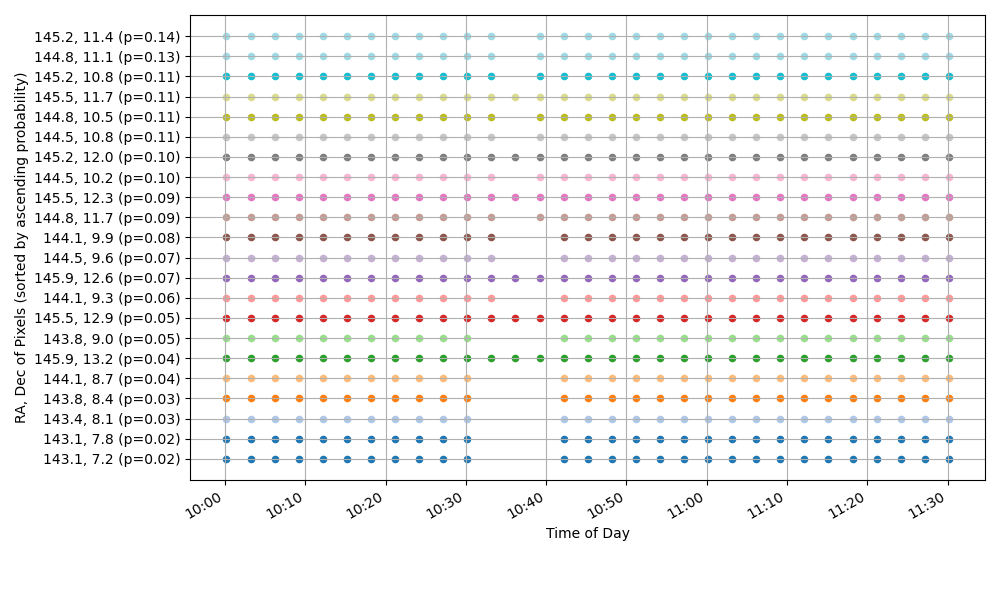}
    \caption{Visibility timeline for each scheduled pointing, expressed in RA/Dec coordinates. Dots indicate availability of pixel (no occultation).}
    \label{fig:visibility}
\end{figure}

Swift orbits Earth approximately every 93.6 minutes. When accounting for Sun and Moon constraints, as well as Earth occultation, a significant portion of the sky becomes temporarily inaccessible during each orbit. In this simulation, only four pointings remain visible throughout the entire 90-minute period. Figure~\ref{fig:occulted} illustrates the HEALPix pixels that are occulted at least once during this interval. Figure~\ref{fig:visibility} shows the final set of pointing coordinates selected by \texttt{tilepy}, highlighting their visibility status throughout the observation window.

Using the 2D-strategy, a total of 22 pointins achieving 74\%coverage of the total probability. The first five pointings corresponding to this coverage are listed in Table~\ref{tab:pgal}. In contrast, for a 3D observational strategy at the event distance of $511 \pm 82$~Mpc, results in much less coverage per pointing struggling to achieve the minimum coverage required and a total coverage of 1\%, as shown in Table~\ref{tab:pgal}. In the case of GW170817 a 3D strategy performs significantly better as shown in Table~\ref{tab:pgal_gw170817}.

\begin{table}[ht]
\centering
\begin{minipage}[t]{0.3\textwidth}
\centering
\begin{tabular}{ccc}
\textbf{RA} & \textbf{Dec} & \textbf{PGW} \\
145.20 & 11.42 & 0.09 \\
144.84 & 11.11 & 0.09 \\
145.20 & 10.81 & 0.07 \\
145.55 & 11.72 & 0.07 \\
144.84 & 10.50 & 0.07 \\
\end{tabular}
\captionof{table}{S250328ae campaign with Swift. Pointings using a 2D strategy. PGW represents the probability covered per pointing independently of the other pointings.}
\label{tab:pgw}
\end{minipage}
\hfill
\begin{minipage}[t]{0.3\textwidth}
\centering
\begin{tabular}{ccc}
\textbf{RA} & \textbf{Dec} & \textbf{PGal} \\
143.43 & 8.69 & 0.005 \\
144.14 & 8.69 & 0.005 \\
 &  &  \\
 &  &  \\
 &  &  \\
\end{tabular}
\captionof{table}{S250328ae campaign with Swift. Pointings using a 3D strategy. PGal represents the galaxy probability covered per pointing independently of the other pointings.}
\label{tab:pgal}
\end{minipage}
\hfill
\begin{minipage}[t]{0.3\textwidth}
\centering
\begin{tabular}{ccc}
\textbf{RA} & \textbf{Dec} & \textbf{PIXFOVPROB} \\
196.88 & -23.64 & 0.17 \\
197.58 & -23.64 & 0.16 \\
197.23 & -23.32 & 0.14 \\
196.88 & -22.99 & 0.14 \\
196.52 & -22.67 & 0.14 \\
\end{tabular}
\captionof{table}{GW170817 campaign with Swift. Pointings from a different event or region. PGal represents the galaxy probability covered per pointing independently of the other pointings.}
\label{tab:pgal_gw170817}
\end{minipage}
\end{table}

\section{Field of view handling}
\begin{figure}
    \centering
    \begin{subfigure}[t]{0.32\textwidth}
        \centering
        \includegraphics[width=\textwidth]{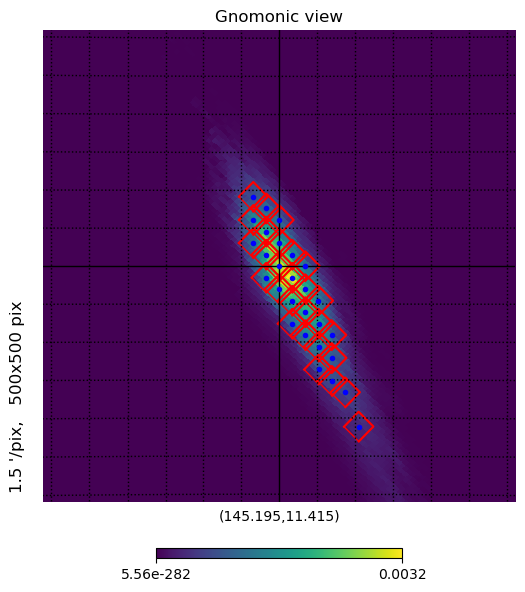}
        \caption{Square FoV (0° rotation)}
    \end{subfigure}
    \hfill
    \begin{subfigure}[t]{0.32\textwidth}
        \centering
        \includegraphics[width=\textwidth]{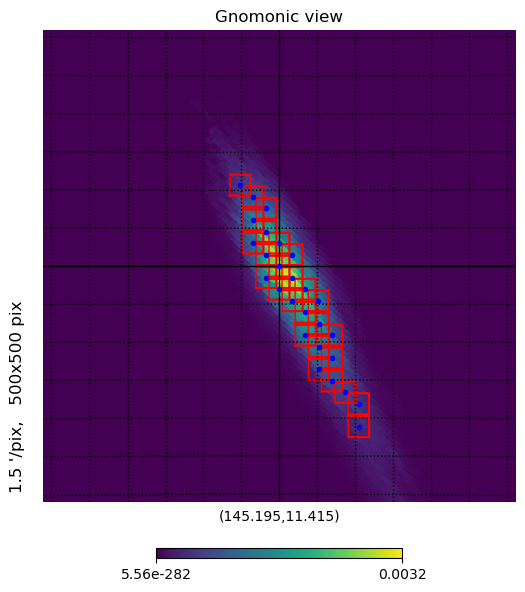}
        \caption{Square FoV (45° rotation)}
    \end{subfigure}
    \hfill
    \begin{subfigure}[t]{0.32\textwidth}
        \centering
        \includegraphics[width=\textwidth]{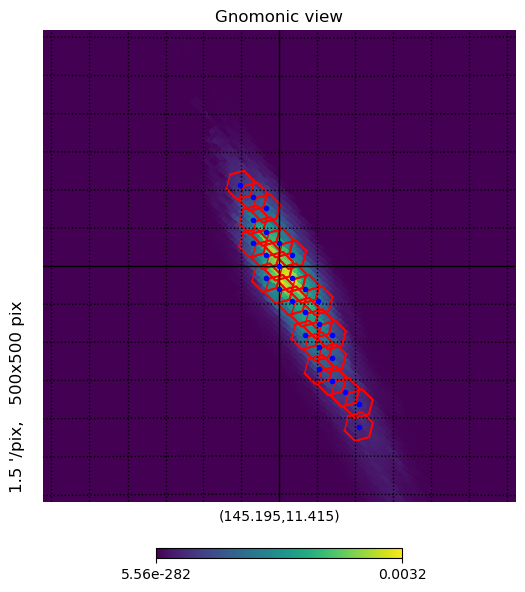}
        \caption{Hexagonal FoV}
    \end{subfigure}
    \caption{Examples of field-of-view handling in \texttt{tilepy}.}
    \label{fig:fov_rotation}
\end{figure}

For orbital instruments, we include support for a variety of field-of-view shapes in \texttt{tilepy}, including circular and symmetric polygonal configurations. These configurations are defined by the user as a number of required sides per polygone (starting 3) Using the \texttt{query\_polygon} function from \texttt{healpy}, we enable the integration of probability within a specified region of the sky. \texttt{tilepy} now supports field-of-view rotation, allowing for accurate modeling of instrument footprints with arbitrary orientations on the sky. This feature is particularly useful for telescopes with non-circular symmetric FoVs, such as rectangular or hexagonal detectors. The user can define the FoV shape as well as the minimum probability coverage per pointing to minimize overlap or maximize coverage. In Figure~\ref{fig:fov_rotation}, we illustrate the impact of rotation handling on sky tiling using three examples with the same configurations in section~\ref{sec:example} and a square FoV (most realistic for Swift-XRT) with no rotation, the same square rotated by $45^\circ$, and a regular hexagonal FoV. Table~\ref{tab:square0}~\ref{tab:square45}~\ref{tab:hexagon} show the first five pointings for these campaigns respectively.

\begin{table}[ht]
\centering

\begin{minipage}[t]{0.30\textwidth}
\centering
\begin{tabular}{ccc}
\textbf{RA} & \textbf{Dec} & \textbf{PGW} \\
145.20 & 11.42 & 0.09 \\
144.84 & 11.11 & 0.09 \\
145.20 & 10.81 & 0.07 \\
145.55 & 11.72 & 0.07 \\
144.84 & 10.50 & 0.07 \\
\end{tabular}
\captionof{table}{S250328ae campaign with Swift. Pointings scheduled with a  a 2D strategy and a square FoV (0° rotation).}
\label{tab:square0}
\end{minipage}
\hfill
\begin{minipage}[t]{0.30\textwidth}
\centering
\begin{tabular}{ccc}
\textbf{RA} & \textbf{Dec} & \textbf{PGW} \\
145.20 & 11.42 & 0.10 \\
144.84 & 11.11 & 0.10 \\
145.20 & 10.81 & 0.08 \\
145.55 & 11.72 & 0.08 \\
144.84 & 10.50 & 0.08 \\
\end{tabular}
\captionof{table}{S250328ae campaign with Swift. Pointings scheduled with a  a 2D strategy and a square FoV (45° rotation).}
\label{tab:square45}
\end{minipage}
\hfill
\begin{minipage}[t]{0.30\textwidth}
\centering
\begin{tabular}{ccc}
\textbf{RA} & \textbf{Dec} & \textbf{PGW} \\
145.20 & 11.42 & 0.12 \\
144.84 & 11.11 & 0.12 \\
145.20 & 10.81 & 0.10 \\
145.55 & 11.72 & 0.10 \\
144.84 & 10.50 & 0.10 \\
\end{tabular}
\captionof{table}{S250328ae campaign with Swift. Pointings scheduled with a  a 2D strategy and a hexagonal FoV.}
\label{tab:hexagon}
\end{minipage}

\end{table}

\section{Observation planner}
\begin{figure}[h!]
    \centering
    \begin{minipage}[t]{0.48\textwidth}
        \centering
        \includegraphics[width=\textwidth]{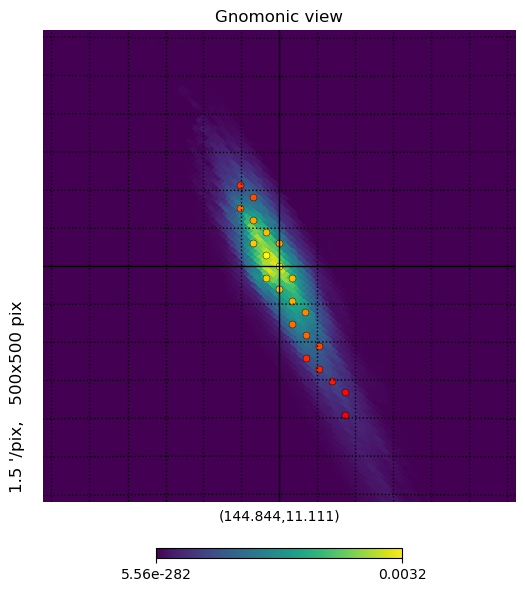}
        \caption{S250328ae campaign with Swift. Initial sky tiling based purely on probability ranking, before applying the observation planner. Colored points indicate the most probable tiles (from yellow to red) without considering slewing constraints.}
        \label{fig:tiling_before}
    \end{minipage}%
    \hfill
    \begin{minipage}[t]{0.48\textwidth}
        \centering
        \includegraphics[width=\textwidth]{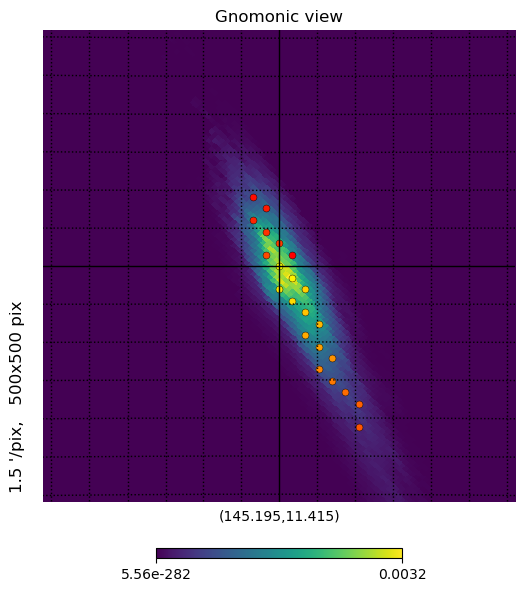}
        \caption{S250328ae campaign with Swift. Optimized sky tiling after applying the observation planner. Colors reflect the observation order (from yellow to red), optimized to reduce slewing while preserving probability coverage.}
        \label{fig:tiling_after}
    \end{minipage}
\end{figure}

In the final stage of the scheduling \texttt{tilepy} provides tools for observation planning like ranking pointings and providing visual aids. We introduce an optional tools to optimize the final schedule based on the user needs. The tool reads a list of candidate fields (RA, Dec) and associated probabilities (e.g., PGW or PGal) and proovides several solutions for pointing clustering. In the example given in Figures~\ref{fig:tiling_before} and~\ref{fig:tiling_after}. We use the S25328ae map with a campaign start date of 2025-07-27 19:30:10, employing a circular field of view (FoV) and relaxing the Earth-limb and Moon avoidance constraints to 1$^\circ$ and 20$^\circ$  respectively. A custom ranking algorithm sorts candidate pointings based on their probability of containing the transient source. Starting from the pointing with the highest probability (PGW), the algorithm iteratively selects the next field that is closest in angular distance to the previously ranked one. This greedy approach is customizable, allowing users to assign more weight to either spatial continuity or high-probability coverage, depending on their observational strategy (e.g., prioritizing continuity to minimize slewing). Other solutions using lightweight AI-based approaches where agglomerative clustering is applied to group spatially adjacent pointings using a configurable angular threshold are provided also. Within each cluster, pointings are ranked by descending probability. The resulting observation plans prioritize high-probability regions while minimizing observatory slewing, making it well suited for efficient and coherent follow-up strategies under operational constraints. \\

\begin{thebibliography}{99}
\setlength{\itemsep}{4pt}

\bibitem{GW170817}
B. P. Abbott, et al. 2017. Phys. Rev. Lett. 119, 161101

\bibitem{GW170817_GRB}
B. P. Abbott, et al. 2017. ApJL, 848 L13 

\bibitem{GW170O817_MWL}
B. P. Abbott, et al. 2017. ApJL, 848, L12

\bibitem{Mochkovitch}
R. Mochkovitch, et al. 2021, A\&A, 651, A83.

\bibitem{Racusin}
J. Racusin, et al. 2017.  arXiv e-prints, 10.48550/arXiv.1708.09292 

\bibitem{Technical_paper}
H. Ashkar, et al. 2021, JCAP, 2021(3), 045

\bibitem{tilepy_2}
M. Seglar-Arroyo, et al. 2024. ApJS 274 1

\bibitem{Patricelli}
B. Patricelli, et al. 2021. PoS, 395, 998

\bibitem{Carossi}
A. Carossi, et al. 2021. PoS, 395, 837. 

\bibitem{saa}
International Geomagnetic Reference Field: the 13th generation, P. Alken, et al. 2021, Earth Planets Space, 73, 49

\bibitem{swift_handbook}
NASA Swift Mission Team, 2022. Swift Technical Handbook, Version 14.0. NASA Goddard Space Flight Center. Available online: \url{https://swift.gsfc.nasa.gov/proposals/tech_appd/swiftta_v14.pdf}

\end{thebibliography}
\end{document}